\input harvmac
\overfullrule=0pt
\def\np{Nucl. Phys.}
\def\pl{Phys. Lett.}


\def\Mk{{\cal M}_k} 
\def\CM{{\cal M}}
\def\Mkl{{\cal M}_{k,l}}

\def\Mbar{{\cal\overline{M}}_1}
\def\Mkbar{{\overline{\cal M}}_k}
\def\xqu{\underline{x}} 
\def\mqu{\underline{m}} 
\def\yqu{\underline{y}} 
\def\Xqu{\underline{X}} 

\def\Pbar{{\bar P}} 
\def\Tr{{\rm Tr}\,}
\def\superF{{\bf F}} 
\def\Ftilde{\widetilde{F}} 
\def\CP{{\cal P}} 


\def\cbar{{\bar c}} 
\def\phibar{\bar \phi} 
\def\Gbar{\bar\Gamma} 


\def\Atilde{\tilde {A}} 
\def\psitilde{\tilde {\psi}} 
\def\phitilde {\tilde {\phi}} 
\def\ctilde {\tilde {c}} 
\def\Dtilde{\widetilde{D}} 
\def\Deltatilde{\widetilde{\Delta}} 

\def\Anew{\tilde{A}^\prime} 
\def\cnew{\tilde{c}^\prime} 
\def\ct{\tilde{c}\,(t)} 
\def\vhat{\widehat{v}} %

\def\BS{L. Baulieu and I.M. Singer, \np\ Proc. Suppl. {\bf B5} (1988) 12.}
\def\BI{ C. Becchi and C. Imbimbo, hep-th/9510003,
\np\ {\bf B462} (1996) 571.}
\def\WITTOP{ E. Witten, \np\ {\bf B 340} (1990) 281.}
\def\DONALDSON{S.K. Donaldson and P.B. Kronheimer, {\it The Geometry of
Four-Manifolds}, (Clarendon Press, Oxford, 1995).}

\def\ANSELMI{D.Anselmi, hep-th/9411049,  \np\ {\bf B 439} (1995) 617.}
\def\ADHM{M.F. Atiyah, N.J. Hitchin, V.G. Drinfeld and Yu.I.
Manin, {\pl}\ {\bf 65A}, (1978) 185.}

{\nopagenumbers
\null
\vskip -2truecm
\abstractfont\hsize=\hstitle\rightline{
\vtop{
\hbox{hep-th/9611113}
\hbox{GEF-TH/96-20}}}
\pageno=0
\vskip 1truecm
\centerline{\titlefont Gauge dependence in topological gauge theories}
\bigskip
\centerline{Carlo M. Becchi\foot{E-mail: becchi@genova.infn.it}}

\centerline{and}

\centerline{ Stefano Giusto\foot{E-mail: giusto@genova.infn.it}}
\vskip 4pt
\centerline{\it Dipartimento di Fisica dell' Universit\`a di Genova}

\centerline{\it Via Dodecaneso 33, I-16146, Genoa, Italy}
\vskip 7pt
\centerline{Camillo Imbimbo\foot{E-mail: imbimbo@infnge.ge.infn.it}}
\vskip 4pt
\centerline{{\it INFN, Sezione di Genova,}}

\centerline{\it Via Dodecaneso 33, I-16146, Genoa, Italy}
\ \medskip
\centerline{ABSTRACT}

We parametrize the gauge-fixing freedom in choosing the Lagrangian of
a topological gauge theory.  We compute the gauge-fixing dependence of
correlators of equivariant operators when the compactified moduli
space has a non-empty boundary and verify that only a subset of these
has a gauge independent meaning.  We analyze in detail a simple
example of such anomalous topological theories, 4D topological
Yang-Mills on the four-sphere and instanton number $k=1$.

\ \vfill
\ifx\answ\bigans
\leftline{GEF-TH/96-20}
\leftline{November 1996}
\else\leftline{November 1996}\fi
\eject}
\ftno=0

\newsec{The problem}

Topological gauge theories \ref\wittop{\WITTOP}\ are characterized by
finite-dimensional gauge orbit spaces --- for example, the moduli space
of Riemann surfaces or the moduli space of (anti)self-dual
four-dimensional euclidean instantons.  Equivariant cohomology classes
of the BRS operator are associated with closed forms on the moduli
space $\CM$ which depend in general on the gauge-fixing choice,
i.e. on the choice of the Lagrangian: different gauge-fixings lead to
closed forms which differ by exact terms.  The evaluation of
correlators reduces to finite-dimensional integration of closed forms
over (cycles of) the moduli space: mathematically this corresponds to
compute intersection numbers on $\CM$.  Thus, only if the integrals of
exact forms over $\CM$ vanish, one is guaranteed {\it a priori} that
the gauge-fixing ambiguity does not affect integrated correlators.

However, in practice, $\CM$ is hardly ever compact. The prototypical
example of the present paper will be topological Yang-Mills theory on
a four-dimensional variety $X$. In this case the non-compactness of
$\Mk$ --- the moduli space of instantons of Pontryagin number $k$ ---
is due to instantons of arbitrarily small size.
Because of this, the proper definition of intersection theory on $\Mk$
\ref\donaldson{\DONALDSON}\ involves the
integration over the compactification $\Mkbar$ of $\Mk$ which includes
instantons of zero-size.  One can show that $\Mkbar$ has no boundary
whenever the ``strata'' $\Mkl$ --- whose points are the $k-$instantons
given by the superposition of a $k-l$-instanton and $l$ zero-size
one-instantons --- are of codimension greater than one, for $l>0$.
This happens when $k$ is in the so-called ``stable range'', which
means that $k$ is sufficiently large ( $4k> 3 b^{+}(X)+4$, when
$G=SU(2)$).  If $k$ belongs to the stable range, integrals of exact
forms over $\Mkbar$ vanish and thus the gauge-fixing ambiguity is
immaterial at the level of physical correlators.

In this article we study topological gauge theories whose compactified
moduli spaces have non-empty boundaries. A simple example
of this situation is topological Yang-Mills on the four-dimensional
sphere $S_4$, gauge group $SU(2)$ and instanton number $k=1$,
not in the stable range: the compactified moduli space is a 5-dimensional
closed ball ${\overline B}_5$, whose boundary, the four-dimensional 
sphere $S_4$, has codimension one. This is also the case when
explicit expressions for the instanton fields are known, so
that a detailed analysis of the gauge-fixing ambiguity is
possible.

We first show that a generic choice for the Lagrangian is
parametrized by a bosonic gauge background $\Atilde(x;m)$ and a
fermionic ghost background $\ctilde (x;m)$.  If the transition
functions for $\Atilde$ and $\ctilde$ are appropriately chosen,
$\Atilde +\ctilde$ defines a connection on a certain
infinite-dimensional bundle on $X\times \Mkbar$; then correlators
of equivariant operators are globally defined closed forms on the
moduli space -- i.e. elements of the De Rham cohomology. If a more
general choice for $\Atilde$ and $\ctilde$ is made, correlators are
not globally defined on $\Mkbar$. They rather have to be interpreted
--- together with their ``descendents'' --- as cocycles of the
\v{C}ech-De Rham sheaf on $\Mkbar$ identified by  a set of local Ward
identities \ref\bi{\BI}.  Well-known results in cohomology theory
ensure that these cocycles are associated with globally defined De
Rham classes. 

Next, we derive the Ward identity that captures the gauge-fixing dependence
of integrated correlators when the moduli space boundary is
non-empty. This will allow us to identify a subset of non-trivial
operators whose correlators have a gauge independent meaning. We shall
see that, for $k=1$ and $X =S_4$, gauge invariant correlators capture
the cohomology of the boundary of the moduli space or, equivalently,
the cohomology of the (compactified) moduli space {\it relative} to
its boundary. We suspect that this is a phenomenon occurring whenever
the compactified moduli space has a boundary, though we do not prove
this in general.

Topological Yang-Mills on $R_4$ for $k=1$ has been considered in a
paper by D. Anselmi \ref\anselmi{\ANSELMI}.  This paper studies the whole 
set of correlators of equivariant closed forms without discussing their
gauge dependence. We shall consider the $R_4$ theory elsewhere. Here
we simply notice that the moduli space of this theory admits a natural
partial compactification into $S_4 \times R_{+}$, which contains a
non-trivial 4-cycle.  Of course, if one restricts oneself to
integration of correlators over this non-trivial 4-cycle, one does not
encounter the boundary-related gauge ambiguities that we discuss here.

\newsec{The Lagrangian}

The BRS transformation laws characterizing the theory are \ref\bs{\BS}:
\eqn\brs{
\eqalign{&s\, A = - D\,c + \psi\qquad s\, \psi\, = -[c , \psi] - D\,\phi\cr
&s\, c \, = - c^2 + \phi \qquad s\, \phi\, = - [c, \phi].\cr}
} 
Here, $s$ is the nilpotent BRS operator and all the fields
are forms on $S_4$ with values in the Lie algebra
of the gauge group $SU(2)$. The one-forms $A = A_\mu d\,x^\mu$  
and $\psi = \psi_\mu d\,x^\mu$ are the gauge connection and the 
gaugino field respectively. The zero-forms  $c$ and $\phi$ 
are the ghost and superghost fields. $A$, $\psi$, $c$, $\phi$
have ghost number 0, 1, 1 and 2 respectively. $D$ is the
covariant exterior differential and commutators are taken in the
Lie algebra of the gauge group. 

The generators of the equivariant cohomology of $s$ satisfy
the descent equations
\eqn\descent{ 
\eqalign{
&s\, \half \Tr F^2 =\, - d\, \Tr F\,\psi\cr
&s\, \Tr F\,\psi =\,  - d\, \Tr \bigl(\phi\, F + \half \psi^2\bigr)\cr
&s\, \Tr\bigl(\phi\, F + \half \psi^2\bigr) =\, -d\,\Tr \phi \psi \cr
&s\, \Tr \phi \psi = -\half d\,\Tr \phi^2 \cr 
&s\, \half\Tr \phi^2 =\, 0,\cr}
}
\noindent where $\Tr$ denotes the invariant Killing form on the Lie algebra
of $SU(2)$.

The Lagrangian of the theory is $s-$trivial, reflecting
the topological character of the theory. We choose it in the
following form:
\eqn\lagraone{
\eqalign{
{\cal L} &=\, s\, \Tr \bigl[\Gbar F_{+} + \cbar\, \Dtilde \star 
(A - \Atilde) + \phibar \Dtilde \star(\psi - {\psitilde}) \cr
&\quad +  x_i W^{(i)}\star (A - \Atilde) \bigr].\cr}
}
We have introduced the anti-fields $\Gbar$, $\cbar$, $\phibar$ 
with values in the gauge group Lie algebra,  
ghost number -1, -1, and -2 and form degree 2, 0, and 0,
respectively. Their BRS transformation properties are:
\eqn\brsantifields{
\eqalign{
&s\, \Gbar =\, - [c ,\Gbar ] + \Lambda \qquad s\, \Lambda =\,  - 
[c ,\Lambda ] - [\phi ,\Gbar] \cr
&s\, \cbar =\, - [c ,\cbar ] + \Sigma \qquad s\, \Sigma =\,  - 
[c ,\Sigma ] - [\phi ,\cbar] \cr
&s\, \phibar =\, - [c ,\phibar ] + \Delta \qquad s\, \Delta =  - 
[c ,\Delta] - [\phi ,\phibar]. \cr}
}
$\Lambda$, $\Sigma$ and $\Delta$ are Lagrangian multipliers
which have the same form degree as $\Gbar$,
$\cbar$ and $\phibar$, respectively.

The Hodge-star duality operator $\star$ acting on forms on $X$
is defined via a space-time background metric $g$. 
$F_+ = \half ( F + \star F)$ is the self-dual part of the curvature
two-form $ F =\, d A + A^2 $. 

$\Atilde = \Atilde (x;m)$ is the gauge field background and $\Dtilde$
is the corresponding covariant derivative.
$m \equiv (m^i)$ with $i=1,\ldots, {\rm dim}\Mk$ labels a point
in the moduli space.
$\Atilde (x;m)$ is a gauge connection which belongs
in the gauge orbit associated
with $m$.

The moduli $m$ are dynamical variables, therefore we extend the action of 
the BRS operator on the moduli space identifying it with
the exterior derivative \bi :
\eqn\brsmoduli{
s\, m^i = d\, m^i \ .}
The last term in  Eq. \lagraone\ fixes the zero modes of $A$ and $\psi$
associated with the tangent vectors to the moduli space.
$\{W^{(i)}\}$ must be
a system of one-forms identifying a basis of the cotangent space to $\Mk$.
$x_i,\ y_i$ are supermultiplets of global Lagrange multipliers:
\eqn\lagrmulti{
s\, x_i =  y_i \ .
}

The basic property of the Lagrangian in Eq. \lagraone\
is that the corresponding functional measure, at fixed moduli,
localizes operators in the algebra generated by $A$, $\psi$,
$c$ and $\phi$ to certain master values. To prove this,  consider first
the following terms in the Lagrangian:
\eqn\mastergauge{
\Tr \bigl[\Lambda F_{+} + \Sigma\, \Dtilde\star (A - \Atilde)
 +  y_i W^{(i)}\star (A - \Atilde) \bigr]\ .
}
\noindent Integrating out 
the Lagrange multipliers $\Lambda$, $\Sigma$ and $y_i$ localizes
$A$ to the background $\Atilde$. 
In fact,  the first term restricts the functional integration
to the anti self-dual connections and the
second term implements the projection to
a local gauge slice which intersects the gauge orbit labelled by
$m$. The integration over $y_i$ projects out connections
which do not belong to the orbit $m$.

Next, turn to the
terms containing the multipliers $\Gbar,\ \cbar$ and $x_i$:
\eqn\masterino{
\Tr \bigl[ \Gbar (\Dtilde\psi)_{+} +\cbar\, \Dtilde\star (\psi
-\Dtilde  c - s\Atilde)
 -  x_i W^{(i)}\star (\psi - \Dtilde  c - s\Atilde)  \bigr].
}
\noindent The first term puts $( \Dtilde \psi)_+ = \bigl(\Dtilde
(\psi - \Dtilde c - s\, \Atilde)\bigr)_+ =0$.  The second and third
terms give the constraint $\psi\  =\, \Dtilde c + s\Atilde$.
Then, the  $\Delta$-dependent term
\eqn\masterdelta{
\Tr \bigl[\Delta\, \Dtilde\star (\psi
-\psitilde )\bigr]=\Tr \bigl[\Delta\, \Dtilde\star
 ( \Dtilde  c +s\Atilde-\psitilde )  \bigr],
}
\noindent leads to the equation
\eqn\masterghost{ \Dtilde\star (\psi - \psitilde) =\, 
\Dtilde \star(s \Atilde + \Dtilde c - \psitilde )=0\ ,
}
\noindent which determines the ghost master field $\ctilde$:
\eqn\ghostmaster{ \ctilde = \star\,  {1\over \Deltatilde}\, \Dtilde\star
( \psitilde - s\, \Atilde),}
\noindent where $\Deltatilde \equiv \star \Dtilde \star \Dtilde 
+ \Dtilde \star \Dtilde \star $ is the Laplacian relative to the background
$\Atilde$. $\Deltatilde$ is not degenerate because of the absence
of reducible  connections on $\Mkbar$.
The same equation \masterghost\ constraints the background
field $\psitilde$ to be equal to the master field for
$\psi$ --- which is $s\, \Atilde + \Dtilde \ctilde$ ---
up to the addition of a one-form $\delta$, with $\Dtilde \star
\delta = \, 0$. However, the background field $\psitilde$ appears in the
Lagrangian \lagraone\ exclusively via the covariant divergence
$\Dtilde \star \psitilde$. Therefore we can identify the master
field for $\psi$ with $\psitilde$ with no loss of generality.

The remaining part  of the Lagrangian is written, after some algebra, 
as:
\eqn\masterphi{
\Tr \bigl[\phibar\, \Dtilde\star\Dtilde (\phi -s\ctilde -\ctilde^2)  
\bigr].
}
Thus the superghost master field $\phitilde $ 
equals $s\, \ctilde + \ctilde^2 $.

Summarizing, we proved the equation 
\eqn\masterfield{
\bigl\langle X\,(A,\, c,\, \psi ,\, \phi\, )\bigr\rangle = X\,(\Atilde, 
\,\ctilde,\, \psitilde ,\, \phitilde ),
}
\noindent with the master fields $\Atilde$, $\psitilde$, $\ctilde$ and
$\phitilde$ obeying BRS transformation rules identical to \brs:
\eqn\brsback{
\eqalign{&s\, \Atilde = - \Dtilde\,\ctilde + \psitilde\qquad 
s\, \psitilde\, = -[\ctilde , \psitilde] - \Dtilde\,\phitilde\cr
&s\, \ctilde \, = - \ctilde^2  + 
\phitilde \qquad\quad s\, \phitilde\, = - [\ctilde, \phitilde].\cr}
}

\newsec{The gauge-fixing dependence of topological field theories}

It follows from the previous analysis that the gauge-fixing freedom in
our Lagrangian corresponds to the choice
of $\Atilde\ (x;m)$ and $\ctilde\ (x;m)$.
We now investigate the dependence of vacuum averages 
of gauge invariant
and equivariant operators on the backgrounds $\Atilde$ and $\ctilde$.
If $\Anew$ is another gauge background, there exists a gauge
transformation $U(x;m) \in SU(2)$, depending in general on $m$,
which relates it to $\Atilde$:
\eqn\changegauge{
\Atilde \rightarrow \Anew = U^{-1} ( \Atilde + v )\, U
}
\noindent where $v \equiv d\,U U^{-1}$ is a space-time one-form with 
values in the Lie algebra of the gauge group. Suppose for a moment that 
we simultaneously change the ghost background as follows
\eqn\changeghost{
\ctilde \rightarrow \cnew = U^{-1} ( \ctilde + \vhat )\, U,
}
\noindent where $\vhat \equiv s\,U U^{-1}$ is a one-form on moduli space
with values in the gauge Lie algebra. It then
follows from Eqs. \brsback\ that the backgrounds $\psitilde$
and $\phitilde$ transform covariantly, i.e. $\psitilde \rightarrow U^{-1}
\psitilde\, U$ and $\phitilde \rightarrow U^{-1}\phitilde\, U$. The BRS
operator in the definition of the Lagrangian \lagraone\ 
transform covariantly under gauge transformations of the quantum fields
acting on the ghost $c$ as in Eq. \changeghost.
Note that these quantum gauge transformations do not coincide
with the classical gauge transformations, which act on
the ghost $c$ homogeneously.  
Thus, if $X$ is a classically 
gauge invariant and {\it equivariant}
operator, its vacuum average is invariant under the simultaneous
variation of the backgrounds in Eqs. \changegauge\ and \changeghost :
\eqn\nochange{
\bigl\langle X\,(A,\, \psi ,\, \phi\, )\bigr\rangle_{\Atilde,\ctilde}
\rightarrow \bigl\langle X\,(A,\, \psi ,\, \phi\, )
\bigr\rangle_{\Anew,\,\cnew} =
\bigl\langle X\,(A,\, \psi ,\, \phi\, )\bigr\rangle_{\Atilde,\,\ctilde}.
}
Eq. \nochange\ shows than an {\it arbitrary} variation of the
bosonic background $\Atilde \rightarrow U^{-1} ( \Atilde + v )\ U$ 
is equivalent to a shift of the ghost background, 
$ \ctilde \rightarrow U^{-1} ( \ctilde - \vhat )\, U$. Therefore
the dependence of the vacuum averages on
the gauge-fixing can be computed by considering their dependence on
arbitrary variations of the ghost background $\ctilde$, keeping
$\Atilde$ fixed. Essentially, this has  already been done in the
context of 2D topological gravity in Ref. \bi . If
\eqn\ghostchange{
\ctilde \rightarrow \cnew \equiv \ctilde + \eta ,
} 
then $\psitilde \rightarrow \psitilde + \Dtilde\eta $. Thus the variation 
of the Lagrangian is 
\eqn\lagrachange{
{\cal L} \rightarrow {\cal L}  - s\,\bigl ( \phibar\, D\star\Dtilde
\ (\cnew -\ctilde)\bigr) =
{\cal L} -  s\, I_\eta {\cal L},
}
\noindent where $I_\eta$ is the operator which shifts the superghost,
$I_\eta\, \phi \equiv \eta$. Therefore:
\eqn\opechange{
\Delta_\eta \langle X \rangle \equiv \langle X\rangle_{\Atilde, \cnew}
-\langle X\rangle_{\Atilde, \ctilde} = \int_0^1\, dt\,\bigl[
s\,\langle I_\eta X\rangle_{\Atilde, \ct } +
\langle I_\eta\,s\, X\rangle_{\Atilde, \ct }\bigr],
}
\noindent where $\ct = t\, \cnew + (1-t)\, \ctilde$ 
interpolates between $\ctilde$ and $\cnew$.

Consider the principal bundle $\CP$ over the moduli space whose 
fiber is the group of local gauge transformations and whose total
space is the space of (anti)-selfdual connections with instanton
number $k$. The choice of the bosonic
background $\Atilde$ corresponds to a choice of a section of this
bundle. In general $\CP$ is non-trivial.
Then, the section $\Atilde$ is only locally defined and one needs to
compare vacuum averages taken with different $\Atilde$ and $\Anew$
related by a gauge transformation $U$ as in Eq. \changegauge . If 
the corresponding ghost backgrounds $\ctilde$ and $\cnew$ 
are related by Eq. \changeghost\ with precisely the {\it same} 
$U$ as in \changegauge, they define a {\it connection} on the principal
bundle $\CP$. Eq. \nochange\ shows that with such a choice of
gauge-fixing on the various patches of moduli space, vacuum averages
of equivariant and gauge invariant operators are globally
defined forms on moduli space. A collection of backgrounds
$(\Atilde, \ctilde)$ satisfying Eqs. \changegauge\ and \changeghost\ 
 is said to define a (global) gauge of De Rham type.
 
However different gauge choices are possible and may be computationally
convenient in certain circumstances. In particular, one can 
choose the gauge-fixing to be of the form $(\Atilde, 0)$
on each patch of the moduli space.  We call this choice
of the (global) gauge of \v{C}ech type for the following
reason. With this gauge choice functional
averages of equivariant observables are not globally 
defined forms; however Eq. \opechange\ shows that in this situation
averages of equivariant observables jump by exact terms when
going from one patch to another. Starting from Eq. \opechange\
it is possible to derive 
a descent of Ward identities whose solution is a cocycle of
the \v{C}ech-De Rham sheaf over the moduli space, equivalent in cohomology
to the global form defined by the De Rham gauge \bi .

In our example, $\Mbar$ is contractible, $\CP$
is trivial and global sections $\Atilde$ exist.  
{\it Any} choice of the ghost background $\ctilde$ defines a 
good connection on $\CP$ and produces
averages of gauge invariant and equivariant
observables $X$ which are globally defined.  If $X$ is $s$-closed,
averages $\langle X\rangle_{\Atilde,
\ctilde}$ computed with different $\ctilde$ differ  
by exact terms, as implied by the Ward identity \opechange .
When $X$ has ghost number five, we can integrate it over 
$\Mbar ={\overline B}_5$,
the closed five-dimensional ball, 
but the result of the integration depends in general 
on the choice of $\ctilde$:
\eqn\changeint{
\Delta_\eta \int_{\Mbar}
\langle X \rangle  = \int_{\Mbar}\int_0^1\, 
dt\, s \,\langle I_\eta X\rangle_{\Atilde, \ct} =
\int_{\partial\Mbar} \int_0^1\, dt \,\langle I_\eta X
\rangle_{\Atilde, \ct}.
}
This equation implies, nonetheless, that $s$-closed operators $X$ which are
{\it independent} of the superghost field $\phi$  have vacuum expectation
values which are independent of the gauge-fixing choice. From
Eq. \descent\ one sees that such $X$ are in the algebra generated
by $\int_{C_3} \Tr F\psi$, for all the space-time 3-cycles
$C_3$. A top form on $\Mbar$ is obtained by considering 
\eqn\topform{
\Omega\ (C_3^{(i)}) = \langle \prod_{i=1}^5 \int_{C_3^{(i)}}\Tr F\psi
\rangle\ .
}
Note that $\Omega$ is
$s-$trivial because the third homology
of $S_4$ is empty:  since $C_3^{(i)} = \partial B_4^{(i)}$, with 
$B_4^{(i)}$ 4-chains in $S_4$, the descent equations \descent\ imply that
\eqn\strivial{
\int_{C_3^{(i)}} \Tr F\psi = \, s\, \int_{B_4^{(i)}} \half \Tr F^2\ .
}
The integral of $\Omega$ over $\Mbar$ is not necessarily zero
because of the non-trivial boundary $\partial\Mbar \equiv S_4$. Since the
pull-back of $\Omega$ on the boundary $\partial \Mbar$ vanishes,
$\Omega$ defines an element of the cohomology of $\Mbar$ relative
to its boundary, which, thanks to the triviality of 
the cohomology of $\Mbar$, coincides
with the cohomology of $\partial\Mbar$. Let us compute this
class explicitely. Consider the bosonic background \ref\adhm{\ADHM}:
\eqn\instanton{
\Atilde = {\cal U}^{-1}\, d\, {\cal U}\ ,
}
\noindent where 
\eqn\quaternion{
{\cal U} = {1\over \sqrt{\rho^{2}+(\xqu-\mqu)^{2}}}
\pmatrix{\rho \cr \xqu - \mqu \cr}\ , 
} 
\noindent is $(2\times 1)$-matrix of quaternions. 
The quaternions $\xqu =x^\mu\, \sigma_\mu$ 
and $\mqu =m^\mu \sigma_\mu$ (with $\sigma_\mu = (1, i\sigma_i)$ where
$\sigma_i, i = 1,2,3$ are the Pauli matrices) correspond to     
the points $(x^\mu)$ and $(m^\mu)$ of $R_4$. 

The coordinates $m$ and $\rho$ appearing in \quaternion\ label
instantons on $R_4$ centered in $m$ with size $\rho$. 
By means of the stereographic projection of the four-sphere
Eq. \quaternion\ defines as well an instanton solution on
$S_4$, taken with a conformally flat metric. 
Thus, $\rho$ and $m$  are
also coordinates on the moduli space of $k=1$ instantons on $S_4$
but only local ones: they  are not to be identified with size and position
of the instantons on $S_4$. To see this, let 
$(y^i)= (y^0, y^\mu)$, with $i=1,\ldots, 5$ and $\mu =1,\ldots,4$,
be cartesian coordinates of $R_5$. 
If one thinks of $\Mbar = {\overline B}_5$ as the unit 5-ball
centered in the origin of $R_5$, $(y^i)$ are global coordinates on
it which are related to $\rho$ and $m$ by means of the equations
\eqn\goodcoordinates{
\eqalign{
&\rho = {\sqrt{ 1 - y^2}\over 1  - y^0} \equiv {\lambda \over 1 - y^0}\cr
& m^\mu = {y^\mu\over 1  - y^0},\cr}
}
\noindent where $y^2 \equiv \sum_i (y^i)^2$. From Eq. \goodcoordinates\
it is clear that $\rho$ and $m$ are good coordinates for $\Mbar$
only on a patch with $y^0 \not= 1$. There is a natural action
of the rotation group $O(5)$ on the cartesian coordinates $(y^i)$, which
induces, via the coordinates transformations \goodcoordinates,
an action on $\rho$ and $m$. This is precisely the $O(5)$ action induced
on the moduli space of instantons by the group of isometries of 
the space-time variety $S_4$. Therefore, the $O(5)$-invariant
$\lambda = \sqrt {1 - y^2} $ in Eq. \goodcoordinates\ is  
to be identified with the size of the instanton and the angular
coordinates of $R_5$ determine the position of its center in $S_4$. 
The center $y^i =0$ of $\Mbar$ represents the $O(5)$-invariant
instanton of maximal size while  
the points on the four-sphere with $\lambda =0$,
which is the boundary of $\Mbar$,  are zero-size instantons.

The $(2\times 1)$-matrix of quaternions in Eq. \quaternion, 
rewritten in terms of the coordinates $(y^i)$, is
\eqn\globalquaternion{
{\cal U} = {1\over \sqrt{\lambda^2 + \Xqu^2}}
\pmatrix{\lambda \cr \Xqu \cr },
} 
\noindent where $\Xqu \equiv (1-y^0)\xqu -\yqu$.

For the evaluation of $\Omega$ the choice of $\ctilde$ is irrelevant,
as shown above. However to illustrate the gauge dependence of the generic
equivariant operators it is useful to consider the following ghost
background
\eqn\superinstanton{
\ctilde = {\cal U}^{-1}\, s\, {\cal U} \ ,
}
\noindent with the same ${\cal U}$ as in Eq. \instanton . 
$\Atilde +\ctilde$ is a connection on the principal bundle over 
$X \times \Mbar = S_4 \times \Mbar$ which is the product of the 
$SU(2)$ bundle
over $S_4$ associated with the instanton and the bundle $\CP$
over $\Mbar$. The curvature of this connection is \bs :
\eqn\supercurvature{
\superF \equiv \Ftilde + \psitilde + \phitilde =  (d + s)(\Atilde +
\ctilde) + (\Atilde +\ctilde)^2\ .
}
The vacuum averages of the operators in the descent equations \descent\ 
are encoded in  the Pontryagin form associated to $\superF$:
\eqn\superdescent{
\half \Tr {\superF}^2 =  \half \Tr {\phitilde}^2 + \Tr \phitilde 
\psitilde + \Tr (\phi\Ftilde + \half \psitilde^2) + \Tr \Ftilde\psitilde + 
\half \Tr \Ftilde^2\ , 
}
\noindent which is a four-form on $S_4\times \Mbar$.  The boundary
of  $S_4\times \Mbar$ is $ S_4\times
\partial\Mbar$. $\partial\Mbar$ is naturally
identified with the space-time 4-sphere, and  we always
imply this identification in the following. 

The computation of the Pontryagin form associated
to $\superF$ parallels the 
evaluation of the Pontryagin form associated to $\Ftilde = 
d\Atilde +\Atilde^2$.  The superconnection $\Atilde + \ctilde$
evaluated at constant $\lambda$ is given by 
\eqn\super{
\Atilde + \ctilde = {\cal U}^{-1} (d + s)\ {\cal U} 
= {1\over \lambda^2 + \Xqu^2}
\bigl[ \Xqu^* (d +s)\ \Xqu - X^\mu (d +s)\ X^\mu\bigr],
}
an expression which is formally identical to the
familiar formula for $\Atilde = {\cal U}^{-1} d\ {\cal U}$.
Analogously, the Pontryagin form associated to $\superF$
pulled-back on the boundary, $\lambda =0$,  is:
\eqn\pullback{
\eqalign{
\half \Tr \superF^2|_{S_4\times\partial\Mbar}& = \, 
\delta (\Xqu)\ \epsilon_{\mu\nu\rho\sigma}
(d+s)X^\mu (d+s)X^\nu (d+s)X^\rho (d+s)X^\sigma \cr
& = \, \delta_{{\cal C}_4}\ - \delta_{{\cal C}_4(\Pbar)}.\cr}
}
\noindent The two four-cycles  
\eqn\fourcycles{
\eqalign{
{\cal C}_4 & = \{ (x,y)\in S_4\times\partial\Mbar\vert\ x=y \}\cr
{\cal C}_4(\Pbar) &= \{ (x,y)\in S_4\times\partial\Mbar\vert\ (x,y) =
(x, \Pbar)  \},\cr}
}
\noindent are the solutions in $S_4\times\partial\Mbar$ of 
the equation $\Xqu =\ 0$.  The point
$\Pbar\equiv (y^0 = 1,\  y^\mu = 0)$ 
is the image in $\partial\Mbar$ of the point in space-time 
defining the stereographic projection.
$\delta_C$ is the delta-function with support on the cycle $C$, 
a form of degree equal to the codimension of $C$.

From Eq. \pullback\ one obtains the pull-back on $\partial
\Mbar$ of the vacuum averages of 
the (non-trivial) operators in the descent, computed in the backgrounds
\instanton\ and \superinstanton:
\eqn\rhamaverages{
\eqalign{
&\langle\ \half\Tr \phi^2 (C_0)\rangle_{\Atilde,\ctilde} = 
\delta_{C_0} -\delta_{\Pbar}\qquad \quad
\langle \int_{C_1} \Tr \phi \psi \rangle_{\Atilde, \ctilde} =
\delta_{C_1}\cr
&\langle \int_{C_2} \Tr (F\  \phi + \half \psi^2) \rangle_{\Atilde, 
\ctilde} =
\delta_{C_2}\qquad
\langle \int_{C_3} \Tr (F\ \psi\rangle_{\Atilde, \ctilde} =
\delta_{C_3}\ .\cr
}
}
\noindent The vacuum averages of the same operators computed in
the gauge $(\Atilde, 0)$ change according to the Ward identity
\changeint :
\eqn\cechaverages{
\eqalign{
&\langle\ \half\Tr \phi^2 (C_0)\rangle_{\Atilde,\ 0} = \,
\langle \int_{C_1} \Tr \phi \psi \rangle_{\Atilde,\ 0} =\, 0\cr
&\langle \int_{C_2} \Tr (F\  \phi + \half \psi^2) \rangle_{\Atilde, 0} =
\, \delta_{C_2} -\ s\,\int_{C_2}\Ftilde \ctilde\qquad \cr
&\langle \int_{C_3} \Tr (F\ \psi\rangle_{\Atilde, 0} =\, 
\delta_{C_3}.\cr
}
}

We see that the class on 
the boundary captured by $\Omega\ (C_3^{(i)})$ depends on the
linking number of the five cycles $C_3^{(i)}$ in $S_4$:
\eqn\relativeclass{
\int_{\Mbar}\Omega\ (C_3^{(i)}) = L (C_3^{(i)}) \equiv  
B_4^{(1)}\cup C_3^{(2)}\cdots\cup C_3^{(5)}\ .
}

Before concluding, let us remark that the
dependence of the vacuum average of $\half \Tr \phi^2$ on
$\Pbar$ reflects its dependence on the choice of ${\cal U}$.
In fact, 
consider a rotation of $O(5)$ acting simultaneously on the space-time
coordinates $x^\mu$ and on the moduli $(y^i)$. It amounts, up to
a gauge transformation, to
multiplying ${\cal U}$ on the left by a moduli-dependent
matrix ${\cal R}$ in ${\rm Sp}(2) \approx O(5)$, which
does not change $\Atilde$ but shifts 
$\ctilde$ by ${\cal U}^{-1}{\cal R}^{-1}(s\ {\cal R})\ {\cal U}$.
The Ward identity \opechange\ then predicts a variation of
$\langle \half \Tr \phi^2\rangle$ which turns out to be equivalent 
precisely to the corresponding $O(5)$ rotation acting on $\Pbar$.
The term $\delta_{\Pbar}$ is absent
in the same correlator computed in the theory on $R_4$, but it
is crucial in our context to guarantee that
\eqn\trivphi{
\int_{\partial\Mbar} \langle\ \half\Tr \phi^2 (C_0)\rangle = \ 0\ .
}

\noindent {\bf Acknowledgements}

It is a pleasure to thank R. Stora for interesting discussions.
This work is partially supported by the ECPR, contract SC1-CT92-0789.

\listrefs
\bye